\documentclass{article}

\usepackage{arxiv}

\usepackage[utf8]{inputenc} % allow utf-8 input
\usepackage[T1]{fontenc}    % use 8-bit T1 fonts
\usepackage{hyperref}       % hyperlinks
\usepackage{url}            % simple URL typesetting
\usepackage{booktabs}       % professional-quality tables
\usepackage{amsfonts}       % blackboard math symbols
\usepackage{nicefrac}       % compact symbols for 1/2, etc.
\usepackage{microtype}      % microtypography
\usepackage{graphicx}
\usepackage{doi}
\usepackage{multirow}%
\usepackage{amsmath,amssymb,amsfonts}%
\usepackage{amsthm}%
\usepackage{mathrsfs}%
\usepackage[title]{appendix}%
\usepackage{textcomp}%
\usepackage{manyfoot}%
\usepackage{algorithm}%
\usepackage{algorithmicx}%
\usepackage{algpseudocode}%
\usepackage{listings}%
\usepackage{booktabs}
\usepackage[table,xcdraw]{xcolor}
\usepackage{multirow}% Required for inserting images
\title{Prediction of Lung Metastasis from Hepatocellular Carcinoma using the SEER Database}

\author{
    Jeff J.H. Kim\thanks{Equal Contribution} \\
    Department of Biomedical Engineering \\
    University of Illinois Chicago \\
    Chicago, IL, USA \\
    \texttt{jkim671@uic.edu} \\
    \And
    George R.~Nahass\footnotemark[1] \\ % Shares the same symbol as Equal Contribution
    Department of Biomedical Engineering \\
    University of Illinois Chicago \\
    Chicago, IL, USA \\
    \texttt{gnahas2@uic.edu} \\
    \And   
    Yang Dai\thanks{Corresponding author} \\
    Department of Biomedical Engineering \\
    University of Illinois Chicago \\
    Chicago, IL, USA \\
    \texttt{yangdai@uic.edu} \\
    \And
    Theja Tulabandhula\footnotemark[2] \\ % Shares the same symbol as Corresponding author
    Department of Information \& Decision Sciences \\
    University of Illinois Chicago \\
    Chicago, IL, USA \\
    \texttt{theja@uic.edu} \\
}

\hypersetup{
pdftitle={A template for the arxiv style},
pdfsubject={q-bio.NC, q-bio.QM},
pdfauthor={David S.~Hippocampus, Elias D.~Striatum},
pdfkeywords={First keyword, Second keyword, More},
}

\begin{document}
\maketitle
% \new
\begin{abstract}

Hepatocellular carcinoma (HCC) is a leading cause of cancer-related mortality, with lung metastases being the most common site of distant spread and significantly worsening prognosis. Despite the growing availability of clinical and demographic data, predictive models for lung metastasis in HCC remain limited in scope and clinical applicability. In this study, we develop and validate an end-to-end machine learning pipeline using data from the Surveillance, Epidemiology, and End Results (SEER) database. We evaluated three machine learning models—Random Forest, XGBoost, and Logistic Regression—alongside a multilayer perceptron (MLP) neural network. Our models achieved high AUROC values and recall, with the Random Forest and MLP models demonstrating the best overall performance (AUROC = 0.82). However, the low precision across models highlights the challenges of accurately predicting positive cases. To address these limitations, we developed a custom loss function incorporating recall optimization, enabling the MLP model to achieve the highest sensitivity. An ensemble approach further improved overall recall by leveraging the strengths of individual models. Feature importance analysis revealed key predictors such as surgery status, tumor staging, and follow-up duration, emphasizing the relevance of clinical interventions and disease progression in metastasis prediction. While this study demonstrates the potential of machine learning for identifying high-risk patients, limitations include reliance on imbalanced datasets, incomplete feature annotations, and the low precision of predictions. Future work should leverage the expanding SEER dataset, improve data imputation techniques, and explore advanced pre-trained models to enhance predictive accuracy and clinical utility.

\end{abstract}

\section{Introduction} \label{intro}

Hepatocellular carcinoma (HCC) ranks as the sixth most common malignancy globally and is a leading cause of cancer-related mortality, with an alarmingly low five-year survival rate of approximately 16\%.\cite{Sung,Sarveazad} Despite advancements in treatment, the prognosis for HCC patients remains grim, particularly when distant metastases develop. 5-year relative survival rates for liver cancer with distant metastasis can go low as 2.5\%. \cite{Gong} Among the common sites for metastasis, the lungs are most frequently affected, accounting for over 70\% of metastatic cases. \cite{Zhan} These metastases significantly diminish survival outcomes, underscoring the need for effective predictive models for quicker treatment.
\newline\newline
The Surveillance, Epidemiology, and End Results (SEER) database, a comprehensive registry of cancer cases in the United States, has been extensively utilized to explore prognostic factors and develop predictive tools for cancer metastases. \cite{hankey1999surveillance} Leveraging machine learning (ML) methodologies, recent studies have successfully created predictive models for HCC metastatic patterns, including lymph node metastases. \cite{chen2022development, sun2024leveraging} 
\newline\newline
However, predictive models tailored specifically for lung metastases in HCC patients remain unexplored. Prior efforts often focus on isolated prognostic factors without integrating advanced computational methods to optimize predictive accuracy.\cite{ye2019risk, shao2023development, zhang2020practical} This gap presents an opportunity to enhance clinical decision-making by identifying high-risk patients earlier, allowing for more targeted surveillance and timely interventions.
\newline\newline
In this study, we aim to develop and validate a robust ML-based predictive model for lung metastasis in HCC patients using the SEER database. By integrating demographic, clinical, and tumor-specific variables, this model seeks to address the limitations of traditional approaches, offering improved precision and clinical utility. Additionally, the findings from this work could contribute to personalized treatment strategies and risk stratification, ultimately improving outcomes for this high-risk population.

\section{Methods} \label{methods}

\subsection{Study Population}
    \begin{figure}
    \centering
    \includegraphics[width=.9\textwidth]{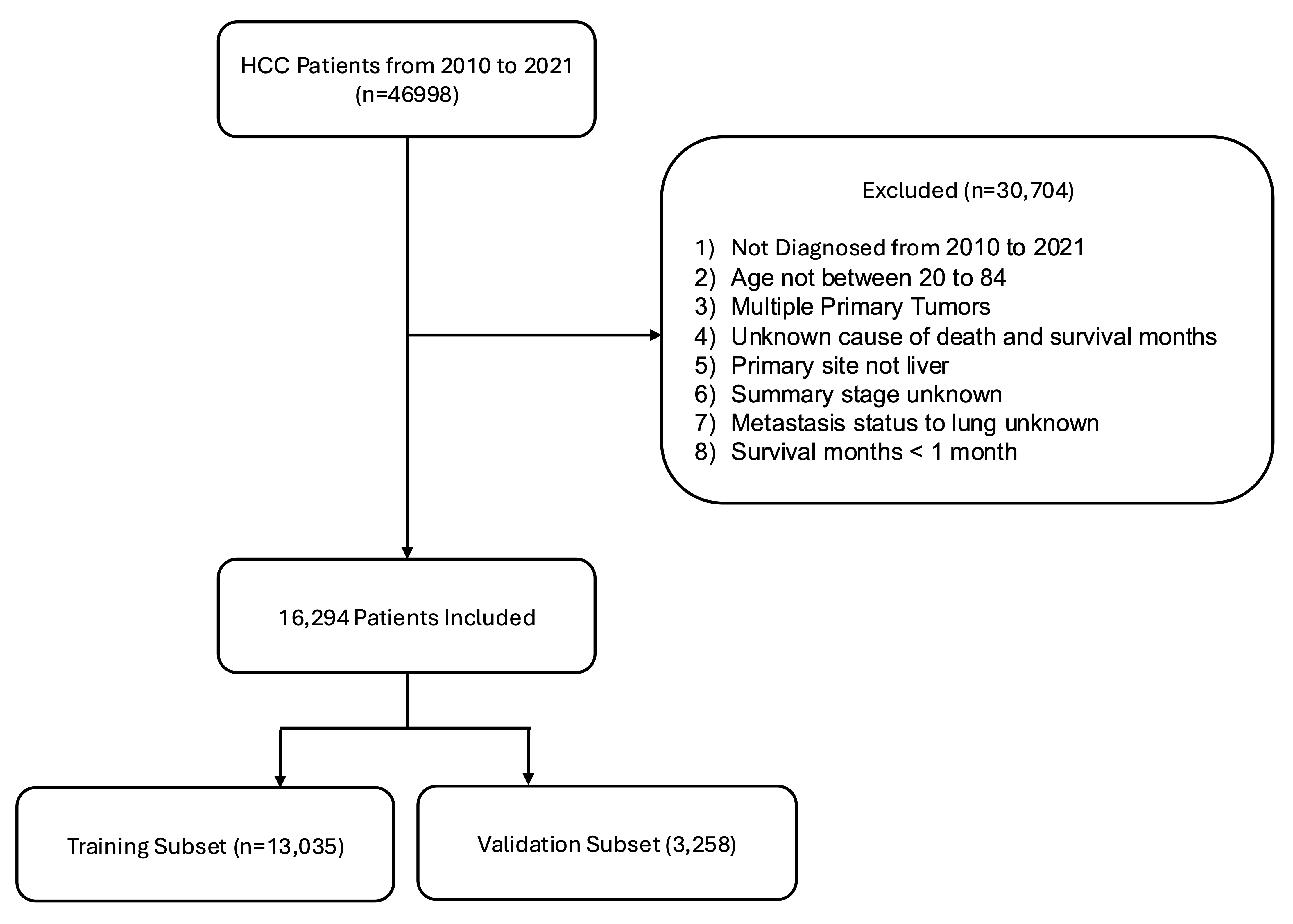}
    \caption{Flowchart illustrating the inclusion and exclusion criteria for the patient population studied, as extracted from the Surveillance, Epidemiology, and End Results (SEER) database.}
\end{figure}

The data is sourced from the Surveillance, Epidemiology, and End Results (SEER) database. \cite{seerdata} The inclusion criteria are as shown in Figure 1: 1) HCC patients diagnosed from 2010 - 2021. 2) age between 20 to 84. 3) Primary site is the liver. 4) One primary cancer only. 5) Known cause of death and followup survival months. 6) Summary and stage known. 7) Known metastasis status to the bone, brain, or lung. 8) Survival months greater than 0. 	

\subsection{Data Preparation}

Redundant features such as the American Joint Committee on Cancer (AJCC) staging grade were consolidated. All feature names were lower cased and non alphanumeric characters were removed. Categorical features were one hot encoded when appropriate, and binary features were mapped to 0 and 1 (e.g. lung\_met). Missing data points were imputed using KNN imputation for numeric features. The minority class was oversampled using the Synthetic Minority Oversampling Technique \cite{Chawla_2002}.

\begin{figure*} [!ht]
    \centering
    \includegraphics[width=.8\textwidth]{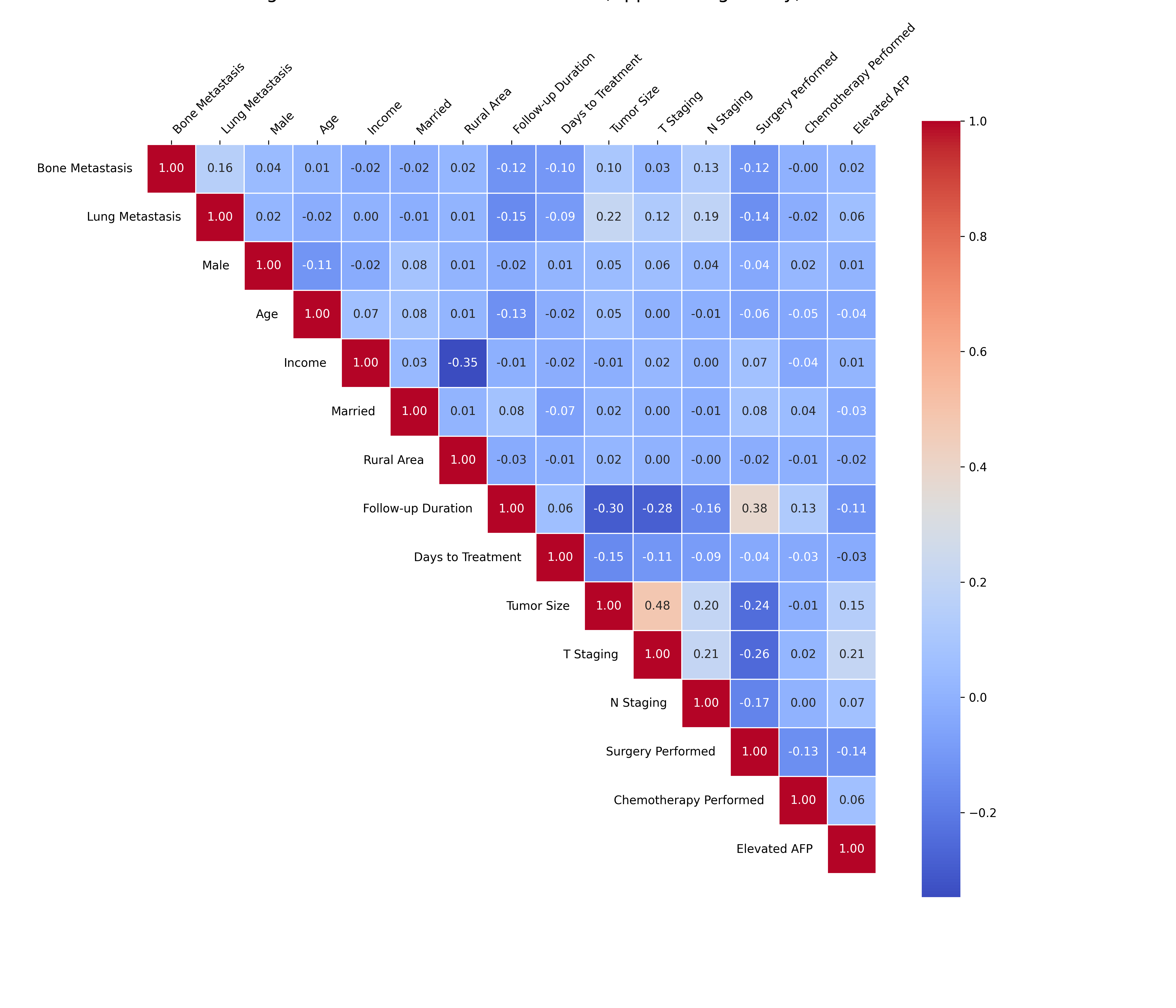}
    \caption{Correlation heatmap displaying the relationships between clinical and demographic variables in the SEER dataset. The color scale represents the strength and direction of the correlation, with red indicating positive correlations and blue indicating negative correlations. The intensity of the color corresponds to the magnitude of the correlation coefficient, with values shown within the cells.}
    \label{fig:corr}
\end{figure*} 

\subsection{Machine Learning}

Three individual machine learning models were trained for binary classification: XGBoost, logistic regression, and random forest. For all machine learning models, optimal hyperparameters were identified via grid search. A train and test split of 80/20 was used in all experiments.  

Feature importances were evaluated for all models. For logistic regression, the absolute values of model coefficients were calculated, as these represent the magnitude of each feature's contribution to the prediction. For XGBoost and random forest models, feature importances were determined using the inherent importance scores provided by the model. These scores were sorted in descending order to identify the top contributors.

\subsubsection{Multi Layer Perceptron}

A three layer multilayer perceptron (MLP) was trained for binary classification. After the first and second hidden a  30\% dropout layer was used. Batch Norm and ReLU was used after every hidden layer. The final output layer consisted of two nodes, and the most probable class was selected following a softmax layer. Softmax was intentionally chosen in the event classification of more classes becomes pertinent in the future. The Adam optimizer with a learning rate of $.001$ was used and the model was trained for $100$ epochs. The model with the highest F1-score was used for downstream tasks.

A combination of the cross entropy loss (\ref{cel}) and recall loss (\ref{rec_loss}) was used for the objective function. In our recall loss formulation, \(\hat{y}_i = \sigma(z_i)\) (where $z_i$ are the output logits) is the predicted positive class, \(y_i\) is the true label, and \(N\) is the total number of samples. 

\begin{equation}\label{rec_loss}
    \mathcal{L}_{\text{recall}} = \frac{1}{N} \sum_{i=1}^{N} (1 - \hat{y}_i) y_i
\end{equation}

\begin{equation}\label{cel}
    \mathcal{L}_{\text{CE}} = -\frac{1}{N} \sum_{i=1}^{N} \left[ y_i \log(\hat{y}_i) + (1 - y_i) \log(1 - \hat{y}_i) \right],
\end{equation}

The cross-entropy loss was weighted according to the distribution of class labels in the training dataset. The recall loss was weighted by a hyperparameter $\gamma$, so the final objective used is:

\begin{equation}\label{fin_loss}
    \mathcal{L}_{\text{Total}} = \mathcal{L}_{\text{CE}} + \gamma \mathcal{L}_{\text{recall}}
\end{equation}

\subsection{Statistical Analysis}

For all machine learning models, the following metrics were evaluated: 
\(\text{Accuracy} = \frac{\text{TP} + \text{TN}}{\text{TP} + \text{TN} + \text{FP} + \text{FN}}\), 
\(\text{Precision} = \frac{\text{TP}}{\text{TP} + \text{FP}}\), 
\(\text{Recall} = \frac{\text{TP}}{\text{TP} + \text{FN}}\), 
\(\text{F1-Score} = 2 \cdot \frac{\text{Precision} \cdot \text{Recall}}{\text{Precision} + \text{Recall}}\), and 
\(\text{AUROC}\), which is the area under the receiver operating characteristic curve.

% {\color{red}Threshold experiments}

\subsection{Hardware}

All experiments were run on an AMD Ryzen 7 7800X3D 8-Core Processor CPU.

\section{Experiments}\label{results}

Below we describe all experiments performed for lung metastasis prediction. 

\subsection{Data Exploration}

We initially probed the SEER dataset to identify any correlations between features (\ref{fig:corr}). Lung metastasis had a positive correlation with tumor size ($r$ = 0.22), T staging ($r$ = 0.12), and N staging ($r$ = 0.19), reflecting the potential role of tumor progression in metastasis.\newline\newline

Follow-up duration exhibited a positive correlation with surgery performed ($r$ = 0.38), suggesting that patients who underwent surgery had longer follow-up periods. Conversely, follow-up duration showed negative correlations with tumor size ($r$ = -0.30), T staging ($r$ = -0.28), and N staging ($r$ = -0.19), indicating that patients with more advanced disease stages or larger tumors tended to have shorter follow-up durations. Income displayed a weak negative correlation with follow-up duration ($r$ = -0.35), suggesting socioeconomic disparities in healthcare access. Other variables, such as surgery performed and elevated AFP, showed minimal direct correlations with metastasis but may have indirect effects captured in the predictive model.

\subsection{Classification}

The following subsections describe the results of the individual performance of all four models evaluated.

\subsubsection{AUROC}
\begin{figure} 
    \centering
    \includegraphics[width=.7\textwidth]{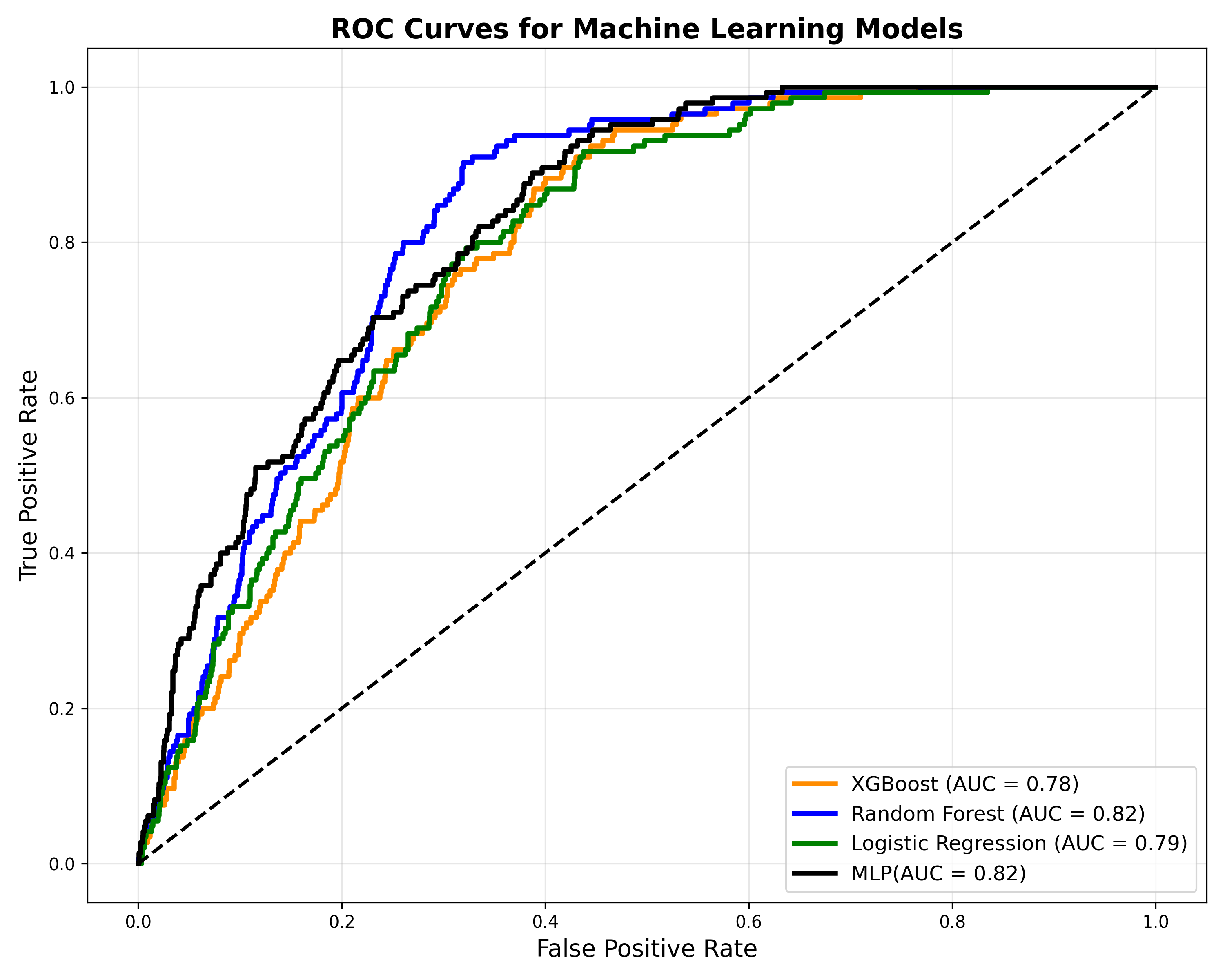}
    \caption{Receiver Operating Characteristic (ROC) curves for machine learning models predicting metastasis status. The curves illustrate the trade-off between true positive rate (sensitivity) and false positive rate for XGBoost, random forest, and logistic regression models.}
    \label{fig:auroc}
\end{figure}

The performance of three machine learning models—XGBoost, logistic regression, and random forest — were evaluated using Receiver Operating Characteristic (ROC) curves (\ref{fig:auroc}). The ROC curves illustrate the trade-off between the true positive rate (sensitivity) and the false positive rate for each model across varying decision thresholds.

The random forest model and MLP models achieved the highest area under the curve (AUC = 0.82), demonstrating the best overall discrimination ability between metastasis and non-metastasis cases. Logistic regression followed with an AUC of 0.79, indicating moderate predictive performance. The XGBoost model had the lowest AUC (0.77) among the three but still performed well above random classification (AUC = 0.50).

\subsection{Confusion Matrices}

The confusion matrices for the XGBoost, logistic regression, and random forest models demonstrate their performance in predicting metastasis status.

\begin{figure}[!hb]
    \centering
    \includegraphics[width=\textwidth]{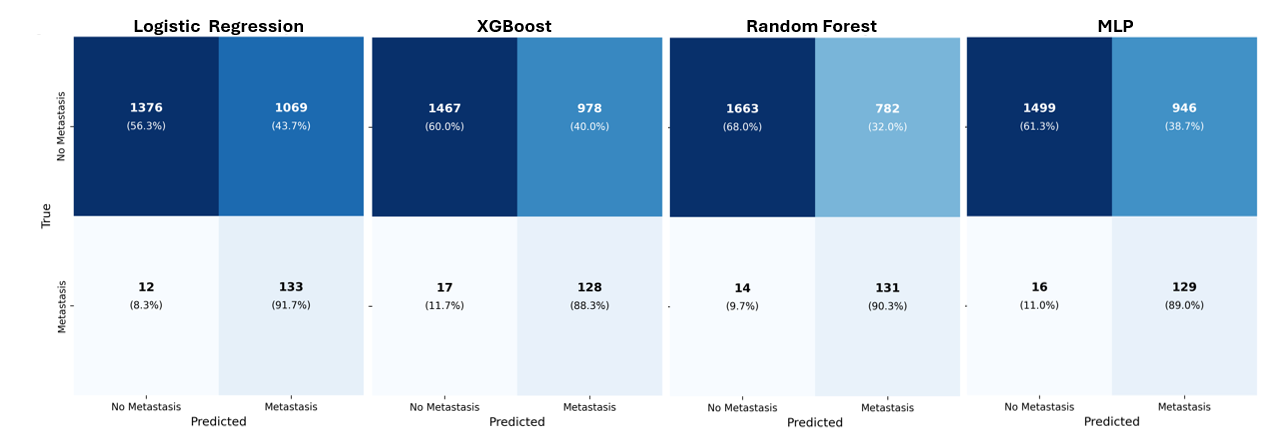}
    \caption{Confusion matrix for logistic regression, XGBoost, random forest, and MLP models on metastasis classification.}
    \label{fig:cms}
\end{figure}

The XGBoost model achieved a high sensitivity for the "Metastasis" class, correctly identifying 128 out of 145 true metastasis cases (88.3\%) with only 17 false negatives (11.7\%) (\ref{fig:cms}). However, its specificity for the "No Metastasis" class was lower, as it misclassified 978 samples as false positives, resulting in a true negative rate of 60.0\%. These results indicate that the XGBoost model is particularly strong at detecting metastasis but at the cost of a higher false-positive rate.

The logistic regression model exhibited an even higher sensitivity for the "Metastasis" class, correctly identifying 133 out of 145 true metastasis cases (91.7\%) with only 12 false negatives (8.3\%) (\ref{fig:cms}. However, the specificity for the "No Metastasis" class was reduced, with 1,069 false positives and a true negative rate of 56.3\%. This indicates that while logistic regression is slightly better at detecting metastasis, it also introduces more false positives compared to XGBoost.

The random forest model outperformed the other two models in terms of specificity, with a true negative rate of 68.0\%, correctly identifying 1,663 non-metastasis cases (\ref{fig:cms}). It also maintained a high sensitivity for the "Metastasis" class, correctly identifying 131 out of 145 true metastasis cases (90.3\%) with only 14 false negatives (9.7\%). These results suggest that the random forest model achieves the best balance between sensitivity and specificity, making it the most robust of the three models. The MLP had a a true negative rate of 61.3\% and True Positive rate of 89\%.

\subsection{Ensembling}

To attempt to leverage the strengths and minimize the weaknesses of any individual models, we created an ensemble of all the individual models trained for lung metastasis classification. In this combined model, if any model predicted a sample as positive the sample was deemed positive. To be classified as negative, four negative predictions were required.

Compared to individual models, the ensemble had the highest recall of 0.77 and an F1 score of 0.24. While the random forest model had the highest individual AUROC and F1-score, its recall was low at 0.5. As such, the ensemble approach, on average, maximized the strengths of the individual models while not being severely worse by any one metric (\ref{tab:ensemble}).

\begin{table}[]
\centering
\resizebox{.8\columnwidth}{!}{%
\begin{tabular}{@{}|c|c|c|c|c|c|c|@{}}
\toprule
 & \textbf{Precision} & \textbf{Recall} & \textbf{Accuracy} & \textbf{AUROC} & \textbf{F1} & \textbf{Average} \\ \midrule
\textbf{XGB}      & 0.15 & 0.26 & 0.87 & 0.78 & 0.19 & 0.45 \\ \midrule
\textbf{Log Reg}  & 0.16 & 0.42 & 0.84 & 0.79 & 0.23 & 0.49 \\ \midrule
\textbf{RF}       & 0.17 & 0.5  & 0.84 & 0.83 & 0.26 & 0.52 \\ \midrule
\textbf{MLP}      & 0.15 & 0.66 & 0.78 & 0.82 & 0.25 & 0.53 \\ \midrule
\textbf{Ensemble} & 0.14 & 0.77 & 0.73 & 0.81 & 0.24 & 0.54 \\ \bottomrule
\end{tabular}%
}
\caption{Machine learning metrics of all individual and ensembled models for metastasis prediction.}
\label{tab:ensemble}
\end{table}

\subsection{Feature Importance}

The XGBoost model highlighted "Surgery Performed" as the most critical feature, followed by "Follow-up Duration" and "Pretreatment AFP Normal" as shown in Figure 5 in supplementary figures (\ref{fig:xg_imp}). Disease staging variables, such as "N Staging," also contributed meaningfully to predictions. Demographics factors, including "Married," "Widowed," and "Urban residency," were among the remaining significant predictors. Interestingly, ethnicity-related variables (e.g., "Korean" and "Other Asian") had minimal yet measurable contributions to the model's performance.

The logistic regression model identified "Pretreatment AFP Normal" as the most important predictor, followed by "Single, Never Married," "N Staging," and "Surgery Performed" (\ref{fig:lr_imp}). Sociodemographic factors, including race (e.g., "Black" and "White") and marital status (e.g., "Married" and "Widowed"), were also significant contributors to the model's predictions. Urban residency and ethnicity further contributed to the model's performance, highlighting the potential influence of demographic variables on metastasis status.

The random forest model emphasized "Follow-up Duration" as the most influential feature, followed by "N Staging" and "Follow-up Year" (\ref{fig:rf_imp}). Tumor-related variables, including "Tumor Size" and "T Staging," also ranked highly in importance, reflecting their direct relevance to cancer progression and metastasis. Other significant features included "Surgery Performed" and "Pretreatment AFP Normal," indicating their roles in treatment outcomes and metastasis risk. Demographic variables such as "Married" and "White" had lower but notable contributions.

To evaluate which features were most important for classification, we compared the ranked feature importances across all three models and compared similar ranks in the top 3, 5 and 10 conserved features. Of the top 3 features for all models, 'surgery not recommended' was ranked by all three models. Of the top 5 features, 'N staging' was also ranked as an important feature. Of the top 10 features, 'Surgery performed, pretreatment AFP Normal, and Follow-up Duration' were also conserved across all three models.

\section{Discussion}\label{disc}

We aimed to develop and validate an end-to-end ML pipeline to predict lung metastasis in HCC patients using the SEER database. Although a prior study by Alkhawaldeh et al. explored the use of machine learning models, such as Random Forest and Easy Ensemble, to predict lung metastasis in HCC patients from the SEER database, their results reported near-perfect performance metrics (AUC = 1, F1 = 0.997). \cite{alkhawaldeh} However, their approach relied extensively on resampling methods, including Easy Ensemble, to address class imbalance. While resampling techniques can improve model performance on imbalanced datasets, they risk introducing biases, such as overfitting, by generating synthetic data that may not represent real-world distributions. This may limit its applicability for broader clinical decision-making or long-term risk prediction.

In contrast, our study aims to provide a robust and reproducible model for predicting lung metastasis in HCC patients, emphasizing real-world applicability and rigorous validation. Unlike prior work, we developed a model grounded in clinically meaningful predictors while ensuring methodological rigor. This approach ensures that the results can be generalized across diverse patient populations and used effectively in clinical decision-making. Four models were developed and used for classification, and, when appropriate, feature-importances were extracted and evaluated. 

While all of the methods we evaluated were able to achieve high AUROC values, indicating that there is good discriminative ability, in many cases, this came at the cost of precision and recall (\ref{tab:ensemble}). However, we have shown that through ensembling where only a single positive vote is required to make a positive classification, we can minimize the weaknesses of any individual model while leveraging the strengths. Of all the metrics evaluated, our ensemble approach achieved the highest average performance with the MLP being the best individual performer.

While it is necessary to evaluate multiple metrics to fairly understand the performance of ML models, in a clinical setting, recall, or the ability to detect true positives, is the most salient. As such, we designed a custom loss function for training the MLP which incorporates both cross entropy loss as well as the recall of the model throughout training. This resulted in the MLP having a recall higher than that of any individual model while still achieving comparable performance in other metrics. 

By only requiring a single positive vote in the ensemble to denote a positive prediction, we subsequently encourage the model to `not miss' positive cases during deployment. As such, while our approach has a very low precision, it may still have utility as a low bandpass filter screening tool. 

Furthermore, we have performed a thorough evaluation into the feature importances predicted by each model. Unsurprisingly, features such as whether or not surgery was performed and the staging of the tumor were conserved across the random forest, XGBoost, and logistic regression model's top 10 most important features. Having an understanding of important features in metastasis prediction is important with respect to the explainability of machine learning models, but also to inform which features should be collected in the clinical pipeline for future model development.  

\subsubsection{Limitations and Future Directions}

Although our approach does provide technical novelty through the introduction of the loss function as well as achieves a high recall and AUROC values, this model would still be unsuitable for wide-spread clinical deployment. The low precision values indicate low degree of confidence in positive predictions, and this inherently limits the utility of such an approach. Datasets with more balanced distributions of the target class would help ensure more robust training, as oversampling, while helpful, will propagate any noise inherent to the original dataset (\cite{alkhawaldeh2023challenges}). Furthermore, improved data imputation methods could enhance the existing noisy dataset. We implement KNN imputation, which has been shown to increase accuracy compared to other imputation techniques, but more sophisticated approaches may yield better results (\cite{aljuaid2016proper}). Furthermore, a significant number of columns (features) in the dataset lacked proper annotations, which limited the model's ability to fully leverage all available information. Improved annotation efforts in the future would provide more comprehensive datasets, enhancing both feature selection and model training.

With the growing number of clinical samples being added to the SEER database, future iterations of this model could benefit from larger and more diverse datasets. Notably, the lung metastasis status data only began being recorded in 2010, so the number of samples will naturally increase over time. This increase in data availability could substantially improve model accuracy and generalizability by enabling better representation of underrepresented cases and reducing the reliance on oversampling methods.

There is also room for improvement in the choice of models used. While we employed relatively simple models in this study, leveraging large pre-trained models, such as transformer-based architectures (e.g., BERT, GPT) or other models pre-trained on biomedical datasets (e.g., ClinicalBERT or BioBERT), could potentially capture complex patterns and relationships in the data. Such advanced models, fine-tuned for specific tasks, have demonstrated substantial performance improvements in biomedical prediction and classification tasks. Future work should explore the integration of these advanced architectures to further enhance the robustness and clinical applicability of the model.

\section{Dataset and Code Availability}

All code is publicly available at \href{https://github.com/monkeygobah/bone-metastasis}{this github repository.}

\bibliographystyle{unsrt}
\bibliography{main}

\section{Supplemental Figures}\label{figs}

\begin{figure}[!h]
\includegraphics[scale=0.5]{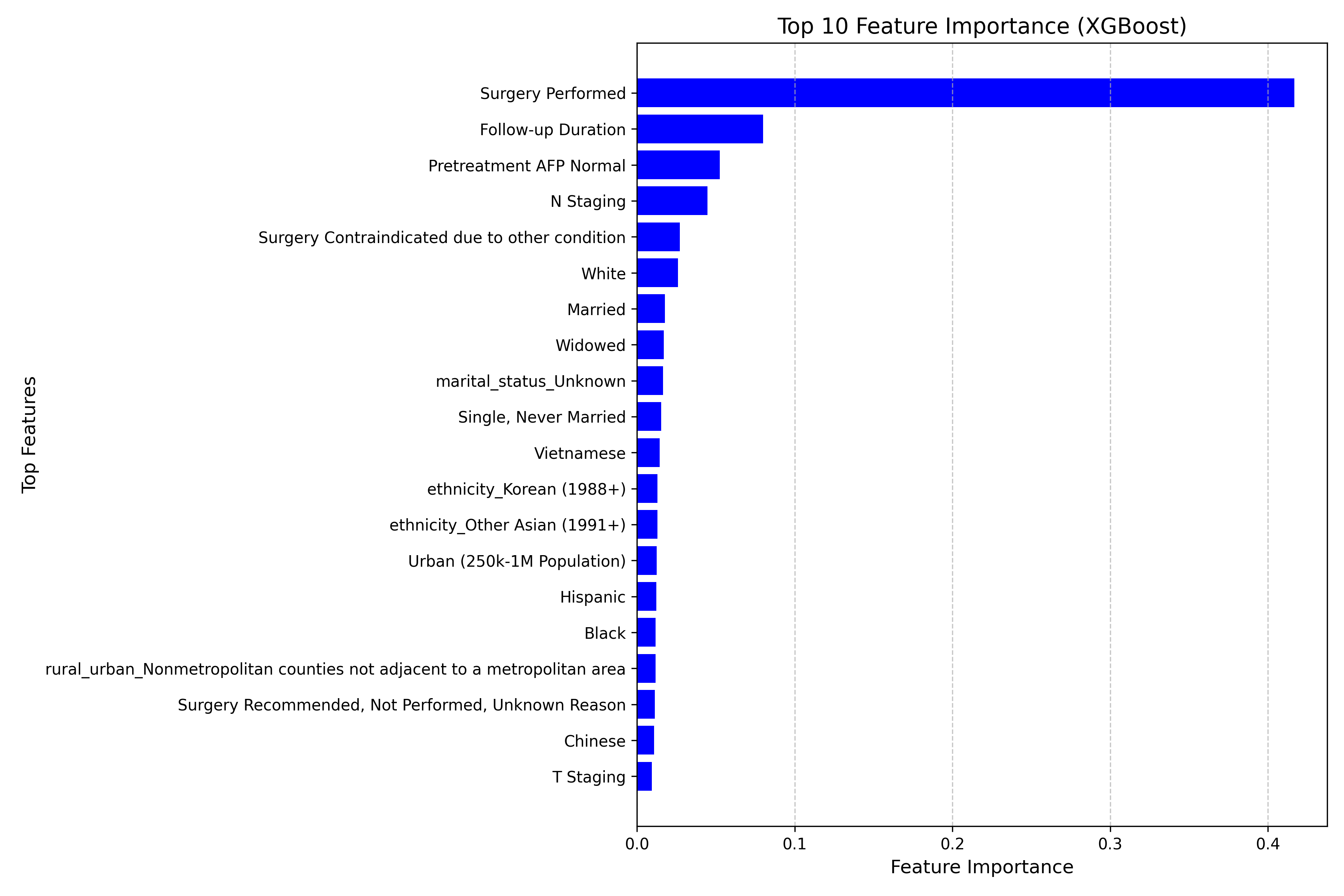}
\caption{Top 20 feature importance scores from the XGBoost model for predicting metastasis status. The most significant features include "Surgery Performed," "Follow-up Duration," and "Pretreatment AFP Normal," which have the highest contributions to model predictions.}
\label{fig:xg_imp}
\end{figure}

\begin{figure}[!h]
\includegraphics[scale=0.5]{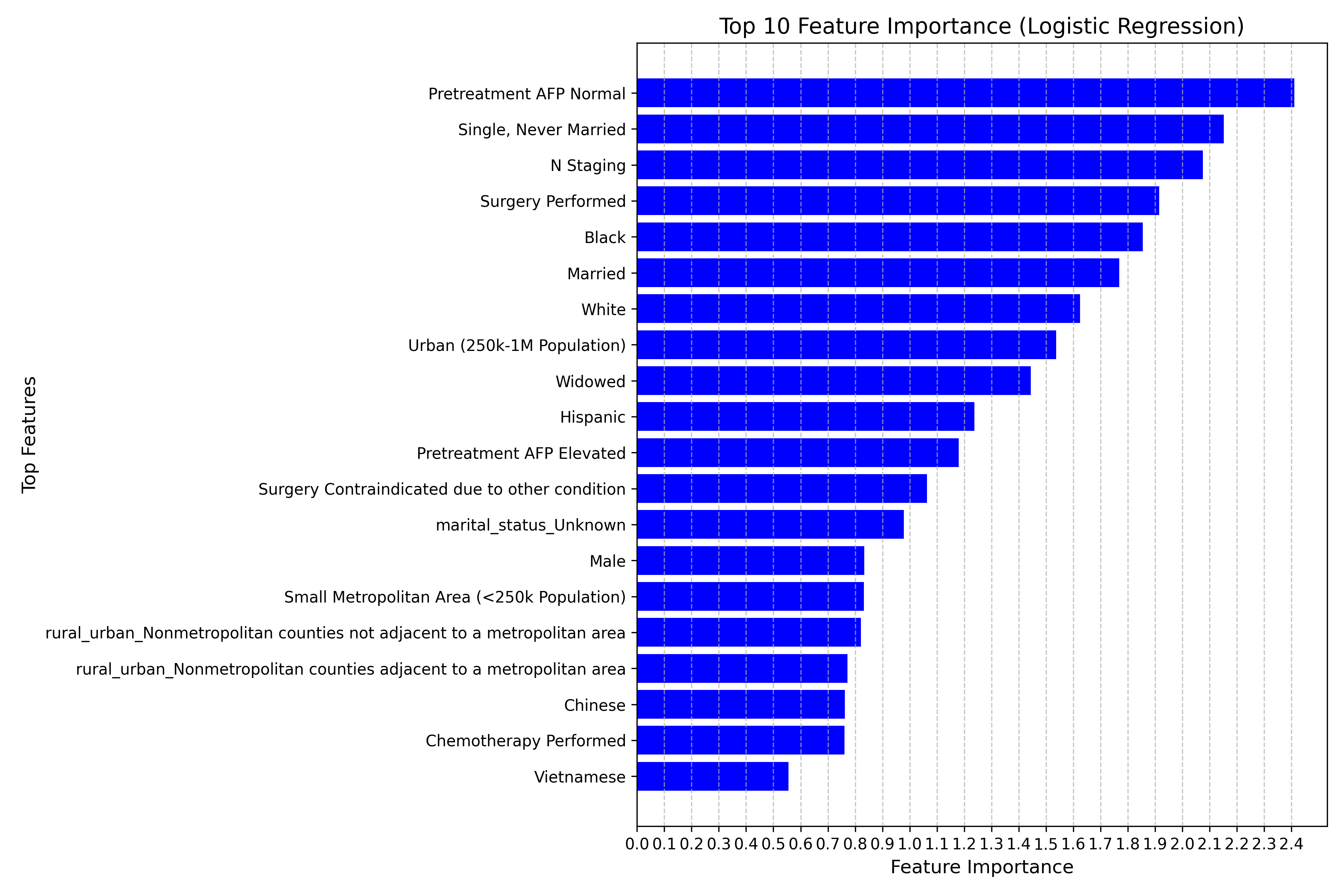}
\caption{Top 20 feature importance scores from the logistic regression model for predicting metastasis status. The most significant features include “Pretreatment AFP Normal”, “Single, Never Married”, and “N Staging" which have the highest contributions to model predictions.}
\label{fig:lr_imp}
\end{figure}

\begin{figure}[!h]
\includegraphics[scale=0.5]{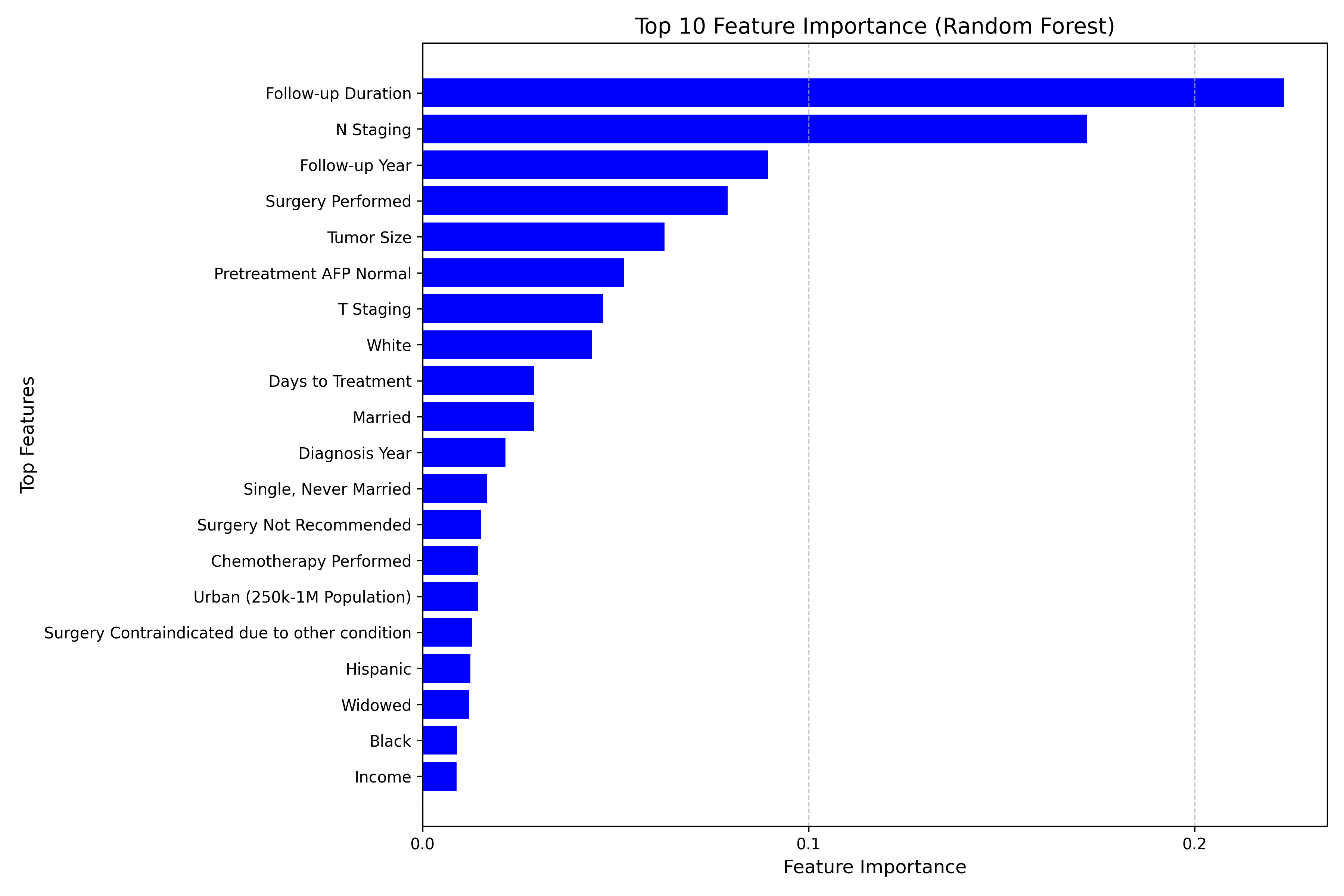}
\caption{Top 20 feature importance scores from the random forest model for predicting metastasis status. The most significant features include “Follow-up Duration”, “N Staging”, and “Follow-up Year“ which have the highest contributions to model predictions.}
\label{fig:rf_imp}
\end{figure}

\end{document}